\documentclass{elsart}

\usepackage {times,amsmath,amssymb,latexsym,graphicx,epic,eepic,verbatim,stmaryrd}
\usepackage[usenames]{color}
\usepackage{multicol}
\usepackage[T1]{fontenc}
\usepackage[latin1]{inputenc}
\usepackage{latexsym,color,verbatim}
\usepackage{amsmath}
\usepackage{amssymb}
\usepackage{amsfonts}
\usepackage{times}
\usepackage{graphicx,lscape}
\usepackage{epsfig}
\usepackage{array}
\usepackage{flafter}
\usepackage[all]{xy}

\begin{document}

\begin{frontmatter}

\title{Notes on Contraction Theory}

\author[lppa]{Nicolas Tabareau}
\author[nsl]{Jean-Jacques Slotine}

\date{}
\corauth{\emph{correspondence to:} 
    N. Tabareau, CNRS, Laboratoire de Physiologie de la Perception et de
    l'Action, Coll{\`e}ge de France, 11 place Marcellin
    Berthelot, 75231 Paris Cedex 05, France, , \emph{Tel.:}
    +33-144271391, \emph{Fax:} +33-144271382 \emph{E-mail:}
    tabareau.nicolas@gmail.com}

\address[lppa]{Laboratoire de Physiologie de la Perception et de l'Action, CNRS
  - Coll{\`e}ge de France,\\
  11 place Marcelin Berthelot, 75231 Paris
  Cedex 05, France.}
\address[nsl]{Nonlinear System Laboratory, Massachusetts Institute of
  Technology,
  \\
  Cambridge, Massachusetts, 02139, USA
}

\begin{abstract}

These notes derive a number of technical results on nonlinear
contraction theory, a comparatively recent tool for system stability
analysis.  In particular, they provide new results on the preservation
of contraction through system combinations, a property of interest in 
modelling biological systems.

\end{abstract}


\end{frontmatter}

\section{Introduction}

Nonlinear contraction theory~\cite{Lohmiller98} is a comparatively
recent tool for system stability analysis. These notes derive a number
of technical results motivated by the theory. In particular, they
provide new results on the preservation of contraction through system
combinations, a property of interest in 
modelling biological systems.

Section 2 analyzes the preservation of contraction through generalized
negative feedback between contracting systems. Section 3 describes a
new system combination, centralized contraction, which also preserves
contraction by aggregation. Section 4 uses standard results from
computer science to simplify the general structure of arbitrary system
combinations, and in particular to exploit intrinsic hierarchical
properties.  Section 5 discusses some applications to nonlinear
attractors, while section 6 describes the estimation of the successive
derivatives of a vector using composite variables.

\section{Negative feedback}

This section analyzes feedback connections which automatically give
rise to contraction with regards to a predefined metric.


Consider two contracting systems, of possibly different dimensions and
metrics, and connect them in feedback, in such a way that the overall
virtual dynamics is of the form
$$
  \frac{d}{dt}  \left( 
  \begin{array}{c}
  \delta {\bf z}_1 \\
  \delta {\bf z}_2 
  \end{array}
  \right) 
  =  \left( 
  \begin{array}{cc}
  {\bf F}_{1} & {-\bf G}({\bf z},t){\bf B} \\
  {\bf G}({\bf z},t)^T {\bf \ A}^T& {\bf F}_{2} 
  \end{array}
  \right) 
  \left( 
  \begin{array}{c}
  \delta {\bf z}_1 \\
  \delta {\bf z}_2 
  \end{array}
  \right) 
$$
with ${\bf A},{\bf B}$ two square matrices. The overall system is contracting if

\begin{enumerate}
\item
  ${\bf A}$ and ${\bf B}$ are symmetric positive definite, and
\item
  there exists $\beta > 0$ such that \\
  $\dot{{\bf A}} + {\bf A}.\bf{F}_1 + {\bf{F}_1}^T  {\bf A} \leq
  -\beta \ A$ \\
  $\dot{{\bf B}} + {\bf B}.\bf{F}_2 + {\bf{F}_2}^T  {\bf B} \leq
  -\beta \ B$
\end{enumerate}

Indeed, we can define the metric
$${\bf M} = \left( 
\begin{array}{cc}
  {\bf A} & {\bf 0} \\
  {\bf 0} & {\bf B} 
\end{array}
\right) $$
We have
$$ \frac{d}{dt}(\delta {\bf z}^T {\bf M} \delta {\bf z}) = 
{\bf M} {\bf F} + {\bf F}^T {\bf M} + \dot{{\bf M}}$$

where the matrix ${\bf M} {\bf F}$ is of the form,
$$ {\bf M} {\bf F} = 
\left(
\begin{array}{cc}
  {\bf A} {\bf F}_{1} & -{\bf A}{\bf G}({\bf z},t){\bf B} \\
  {\bf B}{\bf G}({\bf z},t)^T{\bf A} & {\bf B} .\bf{F}_2
\end{array}
\right) 
$$
Thus
$$ \frac{d}{dt}(\delta {\bf z}^T {\bf M} \delta {\bf z}) =  \delta {\bf z}^T \left( 
\begin{array}{cc}
  \dot{{\bf A}} + {\bf A}  \bf{F}_1 + {\bf{F}_1}^T  {\bf A} & {\bf 0} \\
      {\bf 0} & \dot{{\bf B}} + {\bf B}  \bf{F}_2 + {\bf{F}_2}^T  {\bf B}
\end{array}
\right)  \delta {\bf z}
\leq -\beta \delta {\bf z}^T {\bf M} \delta {\bf z}
$$ 
by hypothesis, which implies that $\delta {\bf z}^T {\bf M} \delta
{\bf z}$ tends exponentially to zero.  Since ${\bf M}$ is positive
definite, this in turn implies that $\delta {\bf z}^T \delta {\bf z}$
tends exponentially to zero.

\section{Centralized contraction}

We extend here the class of combinations of contracting systems
described in \cite{Lohmiller98}. From a practical point of view,
the condition given for feedback combinations may be hard to deal
with, whereas hierarchical combinations are much simpler
but not general enough. 
It is therefore of interest to find a combination having a strong 
expressiveness together with an automatic guarantee of
contraction.

\paragraph*{An almost hierarchical feedback combination.}

\begin{figure} \label{fig:no_loop}
\begin{center}
  \input{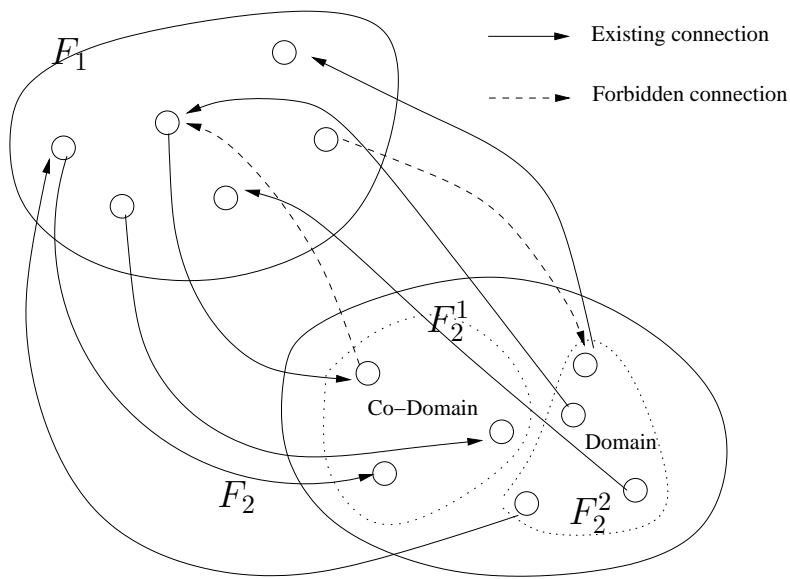}
\caption{A combination that seems to give rise to automatic contraction} 
\end{center}
\end{figure}

The basic idea lies in the following remark.  When looking at the
combination depicted in figure \ref{fig:no_loop}, the loop between
$F_1$ and $F_2$ seems to be illusory, as the domain and co-domain in
$F_2$ are disjoint.  Let's try to formalize this idea.

We consider two contracting system connecting in
feedback in such a way that the second system can be split into ${\bf
  z_2^1}$ and ${\bf z_2^2}$ so to write
$$
\frac{d}{dt} \left (
  \begin{array}{c}
    {\bf \delta z_1} \\
    {\bf \delta z_2^1} \\
    {\bf \delta z_2^2}
  \end{array}
\right )
= 
\left (
  \begin{array}{ccc}
    {\bf F_1} & {\bf G_1}  & \bf 0 \\ 
    \bf 0 & {\bf F_2^{11}}  & {\bf F_2^{12}} \\
    {\bf G_2}  & {\bf F_2^{21}}  & {\bf F_2^{22}}
    \end{array}
    \right )
\left (
  \begin{array}{c}
    {\bf \delta z_1} \\
    {\bf \delta z_2^1} \\
    {\bf \delta z_2^2}
  \end{array}
\right )
$$

Then, a following our idea that this almost represents a hierarchical
combination, we apply the metric 
$$
{\bf \Theta} = \left( 
  \begin{array}{ccc} 
    \bf I & \bf 0 & \bf 0 \\
    \bf 0 & \epsilon^{-1} \bf I & \bf 0 \\
    \bf 0 & \bf 0 & {\epsilon} \bf I
  \end{array}
  \right )
$$

This gives rise to the generalized Jacobian 

$$
\frac{d}{dt} \left (
  \begin{array}{c}
    {\bf \delta z_1} \\
    {\bf \delta z_2^1} \\
    {\bf \delta z_2^2}
  \end{array}
\right )
= 
\left (
  \begin{array}{ccc}
    {\bf F_1} & \epsilon \ {\bf G_1} &  \bf 0\\ 
    \bf 0 & {\bf F_2^{11}}  & \epsilon^{-2} \ {\bf
      F_2^{12}} \\ 
     \epsilon \ {\bf G_2} & \epsilon^{2} \ {\bf F_2^{21}}  & {\bf F_2^{22}}
    \end{array}
    \right )
\left (
  \begin{array}{c}
    {\bf \delta z_1} \\
    {\bf \delta z_2^1} \\
    {\bf \delta z_2^2}
  \end{array}
\right )
$$

This says that, as long as ${\bf G_1}$ and ${\bf G_2}$ are bounded,
they are negligible. However, we have no more guarantee on the
contraction of ${\bf F_2}$ as the matrix of feedback ${\bf F_2^{12}}$
and ${\bf F_2^{21}}$ have been perturbed by $\epsilon$. 

This tells us that we have to restrict this intuition to a particular
kind of feedback within ${\bf F_2}$. 

\subsection{Orientable systems}

The first step is to master the metric used in each local
feedback. Indeed, as in the case of feedback combination, to apply the
combination recursively, we need some guarantees of non interference
between the metric. 
Then idea is to require that the metric use
for each combination only acts on the peripheral system and not on the
centralizer. This leads to the notion of \emph{orientable system}.

\begin{defn}
  A combination between two systems is said to be \emph{orientable}
  if a metric that makes the generalized Jacobian negative definite
  can be written 
  $$
  {\bf M } =
  \left(
    \begin{array}{cc}
      {\bf M'} & {\bf 0} \\
      {\bf 0}  & {\bf I} 
    \end{array}
  \right)
  $$
\end{defn}

\paragraph*{Small gain.}

Consider two contracting systems, of
possibly different dimensions and metrics, and connect them in
feedback, in such a way that the overall virtual dynamics is of the
form 
$$
  \frac{d}{dt}  \left( 
  \begin{array}{c}
  \delta {\bf z}_1 \\
  \delta {\bf z}_2 
  \end{array}
  \right) 
  =  \left( 
  \begin{array}{cc}
  {\bf F}_{1} & {\bf B}{\bf G}({\bf z},t) \\
  {\bf G}({\bf z},t)^T {\bf \ A}^T& {\bf F}_{2} 
  \end{array}
  \right) 
  \left( 
  \begin{array}{c}
  \delta {\bf z}_1 \\
  \delta {\bf z}_2 
  \end{array}
  \right) 
$$
with ${\bf A},\ {\bf B}$ two square matrices.
Note that in this form, ${\bf A}$ and ${\bf B}$ must have the same
dimension. 

Assume now that ${\bf A}$ and ${\bf B}$ satisfy

\begin{itemize}
\item ${\bf B}$ is invertible
\item ${\bf A B}^{-1}$ is constant and symmetric positive definite
\end{itemize}

We can then define the metric $${\bf M} = \left( 
\begin{array}{cc}
  {\bf A}{\bf B}^{-1} & {\bf 0} \\
  {\bf 0} & {\bf I} 
\end{array}
\right) $$  
which can be rewritten ${\bf M} = {\bf \Theta}^T {\bf \Theta}$ with
$${\bf \Theta} = \left( 
\begin{array}{cc}
  \sqrt{{\bf A}{\bf B}^{-1}} & {\bf 0} \\
  {\bf 0} & {\bf I} 
\end{array}
\right) 
$$

Using the fact that 
$$
\sqrt{{\bf A}{\bf B}^{-1}} {\bf B} = (\sqrt{{\bf A}{\bf B}^{-1}})^{-1}
{\bf A} 
$$
we have 
$$
\left( 
  \begin{array}{cc}
  \sqrt{{\bf A}{\bf B}^{-1}} \bf{F}_1 (\sqrt{{\bf A}{\bf
      B}^{-1}})^{-1}  & \sqrt{{\bf A}{\bf B}^{-1}} {\bf B} {\bf
    G}({\bf z},t)  \\ 
  {\bf G}({\bf z},t)^T (\sqrt{{\bf A}{\bf B}^{-1}} {\bf B})^T & \bf{F}_2  
\end{array}
\right) 
$$

Applying a standard result for small gain feedback (see
\cite{Slotine03}), we can conclude on the contraction of the system
if $\sqrt{{\bf A}{\bf B}^{-1}} \bf{F}_1 (\sqrt{{\bf A}{\bf
      B}^{-1}})^{-1}$ is negative definite and the
following inequality holds
\begin{equation} \label{ineq:small gain}
\sigma^2(\sqrt{{\bf A}{\bf B}^{-1}} {\bf B} {\bf G}({\bf z},t)) <
\lambda((\sqrt{{\bf A}{\bf B}^{-1}} \bf{F}_1 (\sqrt{{\bf A}{\bf
      B}^{-1}})^{-1})_s) \lambda((\bf{F}_2)_s)
\end{equation}

We can conclude that this system is an orientable scaling-robust
system.

Note that the above assumptions on ${\bf A}$ and ${\bf B}$ are verified in the
common case that ${\bf A}$ is constant and symmetric positive definite and
${\bf B} = \lambda {\bf I}$ with constant $\lambda > 0$.

\paragraph*{Negative feedback.}

In the same way, for a system of the form
$$
  \frac{d}{dt}  \left( 
  \begin{array}{c}
  \delta {\bf z}_1 \\
  \delta {\bf z}_2 
  \end{array}
  \right) 
  =  \left( 
  \begin{array}{cc}
  {\bf F}_{1} & -{\bf B}{\bf G}({\bf z},t) \\
  {\bf G}({\bf z},t)^T {\bf \ A}^T& {\bf F}_{2} 
  \end{array}
  \right) 
  \left( 
  \begin{array}{c}
  \delta {\bf z}_1 \\
  \delta {\bf z}_2 
  \end{array}
  \right) 
$$
we can construct a orientable metric such that the system is
contracting if $\sqrt{{\bf A}{\bf B}^{-1}} \bf{F}_1 (\sqrt{{\bf
    A}{\bf B}^{-1}})^{-1}$ is negative definite.

\subsection{Orientable scaling-robust systems}

\begin{defn}
  A combination between two systems is said to be \emph{orientable
    scaling-robust} if transforming the system by the metric $D_\epsilon =  
  \left (
    \begin{array}{cr}
      \bf I & \ \bf 0\\
      \bf 0 & \epsilon \ \bf I
    \end{array}
  \right )$ ($\epsilon >0$)
  leads to an orientable combination with metric   
  $\left (
    \begin{array}{cr}
      M_\epsilon & \bf 0\\
      \bf 0 & \bf I
    \end{array}
  \right )$.
  We further require that $M_\epsilon \xrightarrow{\epsilon
    \rightarrow 0} 0$.  
\end{defn}

\begin{rem}
  The definition above says that the dynamics
  $$
  \frac{d}{dt} \left (
    \begin{array}{c}
      {\bf \delta x^1} \\
      {\bf \delta x^2}
    \end{array}
  \right )
  = 
  \left (
    \begin{array}{lr}
      {\bf F^{11}}  & \epsilon^{-1} \ {\bf F^{12}} \\ 
      \epsilon \ {\bf F^{21}}  & {\bf F^{22}}
    \end{array}
  \right )
  \left (
    \begin{array}{c}
      {\bf \delta x^1} \\
      {\bf \delta x^2}
    \end{array}
  \right )
  $$
  are orientable for all $\epsilon$.
\end{rem}

\paragraph*{Small gain and negative feedback.}

Let us come back to the two previous examples.
It is clear that applying the metric 
${\bf D}_\epsilon =  
  \left (
    \begin{array}{cr}
      \bf I & \ \bf 0\\
      \bf 0 & \ \ \epsilon \ \bf I
    \end{array}
  \right )$
for $\epsilon >0$ leads to another contracting system with metric
$${\bf M}_\epsilon = \left( 
\begin{array}{cc}
  \epsilon \ {\bf A}{\bf B}^{-1} & {\bf 0} \\
  {\bf 0} & {\bf I} 
\end{array}
\right) $$  
So an orientable small gain (resp. negative feedback) is automatically
scaling-robust.

\subsection{Centralized contraction}

\begin{figure} \label{fig:star_loop}
\begin{center}
  \input{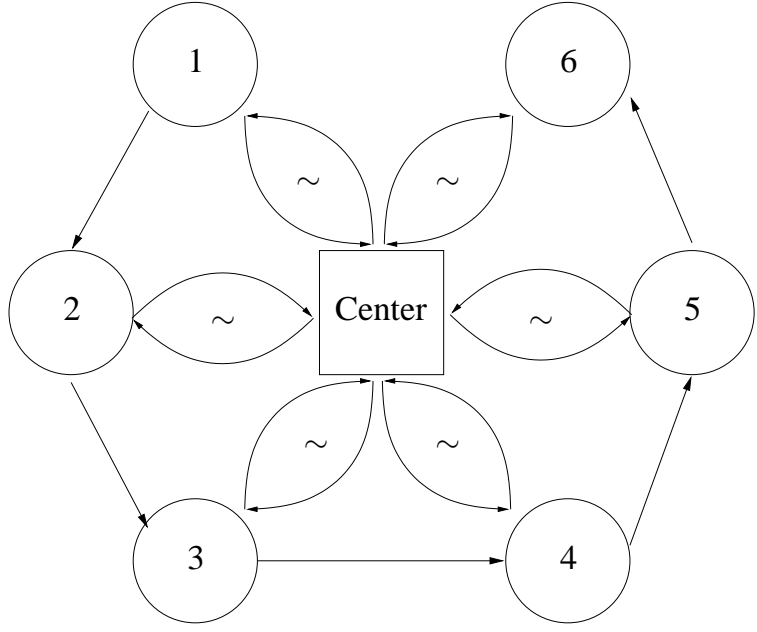}
\caption{A centralized combination of contracting systems} 
\end{center}
\end{figure}

Assume that we have, as in figure \ref{fig:star_loop}, $n$ systems
connecting to a particular system called the \emph{center} in such a way that
every connection to the center is \emph{orientable} and
\emph{scaling-robust}. Assume also that the connection between
the different peripheral systems is hierarchical.

That kind of system can be rewritten 

$$
\frac{d}{dt} \left(
\begin{array}{c}
  \delta {\bf z_6} \\ \delta {\bf z_5} \\ \delta {\bf z_4} \\ \delta
  {\bf z_3} \\ \delta {\bf z_2} \\ \delta {\bf z_1} \\ \delta {\bf
  z_{C}} \\  
\end{array}
\right)
=
\left(
\begin{array}{ccccccc}
  \bf F_6 & \bf X & \bf X & \bf X & \bf X & \bf X & \bf F_6^2 \\
  \bf 0 & \bf F_5 & \bf X & \bf X & \bf X & \bf X & \bf F_5^2 \\
  \bf 0 & \bf 0 & \bf F_4 & \bf X & \bf X & \bf X & \bf F_4^2 \\
  \bf 0 & \bf 0 & \bf 0 & \bf F_3 & \bf X & \bf X & \bf F_3^2 \\
  \bf 0 & \bf 0 & \bf 0 & \bf 0 & \bf F_2 & \bf X & \bf F_2^2 \\
  \bf 0 & \bf 0 & \bf 0 & \bf 0 & \bf 0 & \bf F_1 & \bf F_1^2 \\
  \bf F_6^1 & \bf F_5^1 & \bf F_4^1 & \bf F_3^1 & \bf F_2^1 & \bf F_1^1 & \bf C
\end{array}
\right)
\left(
\begin{array}{c}
  \bf \delta z_6 \\ \bf \delta z_5 \\ \bf \delta z_4 \\ \bf \delta z_3
  \\ \bf \delta z_2 \\ \bf \delta z_1 \\ \bf \delta z_{C}
\end{array}
\right)
$$

For that particular virtual system, let us use the metric
$$
\left(
\begin{array}{ccccccc}
  \bf I & \bf 0 & \bf 0 & \bf 0 & \bf 0 & \bf 0 & \bf 0 \\
  \bf 0 & \bf \epsilon^{-1} I & \bf 0 & \bf 0 & \bf 0 & \bf 0 & \bf 0 \\
  \bf 0 & \bf 0 & \bf \epsilon^{-2} I & \bf 0 & \bf 0 & \bf 0 & \bf 0 \\
  \bf 0 & \bf 0 & \bf 0 & \bf \epsilon^{-3} I & \bf 0 & \bf 0 & \bf 0 \\
  \bf 0 & \bf 0 & \bf 0 & \bf 0 & \bf \epsilon^{-4} I & \bf 0 & \bf 0 \\
  \bf 0 & \bf 0 & \bf 0 & \bf 0 & \bf 0 & \bf \epsilon^{-5} I & \bf 0 \\
  \bf 0 & \bf 0 & \bf 0 & \bf 0 & \bf 0 & \bf 0 & \bf \epsilon I
\end{array}
\right)
$$ 

which leads to the system

$$
\frac{d}{dt} \left(
\begin{array}{c}
   \bf \delta z_6 \\ \bf \delta z_5 \\ \bf \delta z_4 \\ \bf \delta
  z_3 \\ \bf \delta z_2 \\ \bf \delta z_1 \\ \bf \delta z_{C}
\end{array}
\right)
=
\left(
\begin{array}{ccccccc}
  \bf F_6 & \bf \epsilon X & \bf \epsilon^2 X & \bf \epsilon^3 X & \bf
  \epsilon^4 X & \bf  \epsilon^5X & \bf \epsilon^{-1} F_6^2 \\ 
  \bf 0 & \bf F_5 & \bf \epsilon X & \bf \epsilon^2 X & \bf \epsilon^3 X &
  \bf \epsilon^4 X & \bf \epsilon^{-2} F_5^2 \\ 
  \bf 0 & \bf 0 & \bf F_4 & \bf \epsilon X & \bf \epsilon^2 X & \bf
  \epsilon^3 X & \bf \epsilon^{-3} F_4^2 \\ 
  \bf 0 & \bf 0 & \bf 0 & \bf F_3 & \bf \epsilon X & \bf \epsilon^2 X
  & \bf \epsilon^{-4} F_3^2 \\ 
  \bf 0 & \bf 0 & \bf 0 & \bf 0 & \bf F_2 & \bf \epsilon X & \bf
  \epsilon^{-5} F_2^2 \\ 
  \bf 0 & \bf 0 & \bf 0 & \bf 0 & \bf 0 & \bf F_1 & \bf \epsilon^{-6} F_1^2 \\
  \bf \epsilon F_6^1 & \bf \epsilon^2 F_5^1 & \bf \epsilon^3 F_4^1 & \bf
  \epsilon^4 F_3^1 & \bf \epsilon^5 F_2^1 & \bf \epsilon^6 F_1^1 & \bf C 
\end{array}
\right)
\left(
\begin{array}{c}
  \bf \delta z_6 \\ \bf \delta z_5 \\ \bf \delta z_4 \\ \bf \delta z_3
  \\ \bf \delta z_2 \\ \bf \delta z_1 \\ \bf \delta z_{C}
\end{array}
\right)
$$

Since all the couple $(F_i^1,F_i^2)$ are assumed to be orientable scaling-robust,
for any small $\epsilon$, there is a constant metric ${\bf M}$ such
that the symmetric part of the
generalized Jacobian can be written, when $\epsilon$ tends to zero, as 
$$
\left(
\begin{array}{ccccccc}
  \bf H_6 & \bf 0 & \bf 0 & \bf 0 & \bf 0 & \bf 0 & \bf K_6 \\
  \bf 0 & \bf H_5 & \bf 0 & \bf 0 & \bf 0 & \bf 0 & \bf K_5 \\
  \bf 0 & \bf 0 & \bf H_4 & \bf 0 & \bf 0 & \bf 0 & \bf K_4 \\
  \bf 0 & \bf 0 & \bf 0 & \bf H_3 & \bf 0 & \bf 0 & \bf K_3 \\
  \bf 0 & \bf 0 & \bf 0 & \bf 0 & \bf H_2 & \bf 0 & \bf K_2 \\
  \bf 0 & \bf 0 & \bf 0 & \bf 0 & \bf 0 & \bf H_1 & \bf K_1 \\
  \bf K_6^T & \bf K_5^T & \bf K_4^T & \bf K_3^T & \bf K_2^T & \bf
  K_1^T & \bf C  
\end{array}
\right)
$$
where the matrices 
$$
\left(
\begin{array}{cc}
  \bf H_i & \bf K_i \\
  \bf K_i^T & \bf C 
\end{array}
\right)
$$
are all negative definite. 

Applying a basic result of matrix analysis thus yields the condition
$$
{\bf C} < \sum_i {\bf K_i^T H_i^{-1} K_i}
$$
which is equivalent to
$$
\sum_i \lambda({\bf K_i^T H_i^{-1} K_i C^{-1}}) < 1
$$
A sufficient condition is thus
$$
\sum_i \sigma({\bf K_i})^2\lambda({\bf H_i})^{-1} < \lambda({\bf C})
$$
or even the less general but easier to verify inequality
$$
\sum_i \sigma({\bf K_i})^2 < \lambda({\bf C})\min_i\lambda({\bf H_i})
$$

We call this case \emph{centralized contraction}.

\subsection{Going further}

The structure of centralized contraction is a general scheme that can
be extended to more complex structures. We present here two basic
extensions.

\paragraph*{Multiple layers}

\begin{figure}[h!]
\begin{center}
  \includegraphics[scale=0.8]{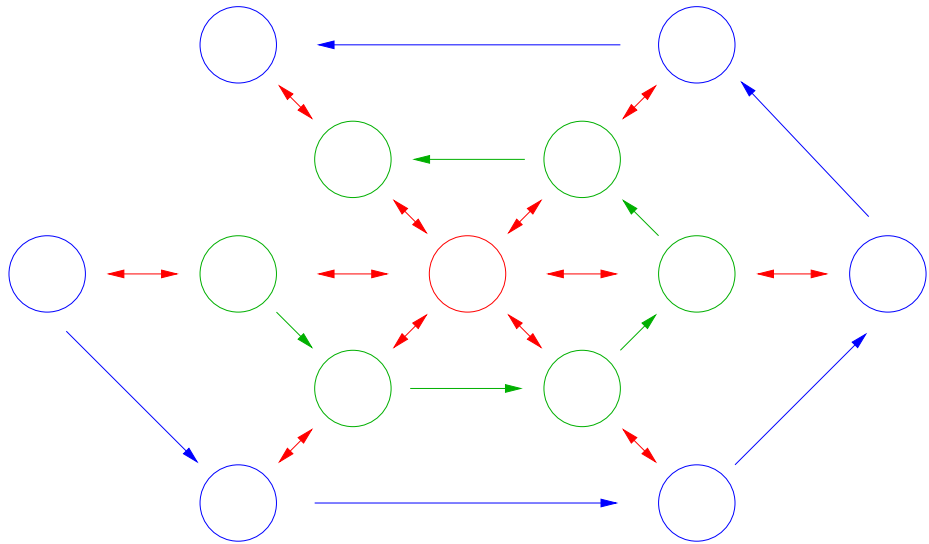}
  \caption{Centralized contraction for multiple layers}
  \label{fig:multiple layers}
\end{center}
\end{figure}

We can apply the result of centralized contraction even if the
peripheral system is composed of different layers. For example, the
system described in figure \ref{fig:multiple layers} is automatically
contracting providing that the red connections are orientable scaling
robust.  

\paragraph*{Multiple centers}

In biological systems, it is often of interest to consider a
centralizer which is composed of multiple systems. In that case, the
contraction can be guaranteed if all connections to the center
considered as a whole are orientable scaling robust.

\section{Strongly connected components}

In computational neuroscience as in many biological fields, we have to
deals with large systems.  Here we exploit a standard
algorithm from computer science~\cite{Knu} to decompose a large system
into sub-systems, in a such way that the contraction of the overall
system can be deduced from the contraction of the smaller sub-systems.

\begin{defn}
A \emph{strongly connected component} of a directed graph $G=(V,E)$ is
a maximal set of vertices $U \subset V$ such that 
for all $u,v \in U$, $u$ is reachable from $v$ and
$v$ is reachable from $u$.
\end{defn}

\begin{prop}
Any directed graph is a union of \emph{strongly connected components}
plus edges to join the components together.
\end{prop}

Thus, we are able to distinguish between micro-systems which are 
connected in feedback combination or not. Indeed, we can state 
the proposition:

\begin{prop}
Two sub-systems of large systems are in feedback combination iff
those two systems belongs to the same strongly connected component.
\end{prop}

Let us now describe the algorithm to compute such a decomposition.

\begin{figure}
\begin{center}

\includegraphics{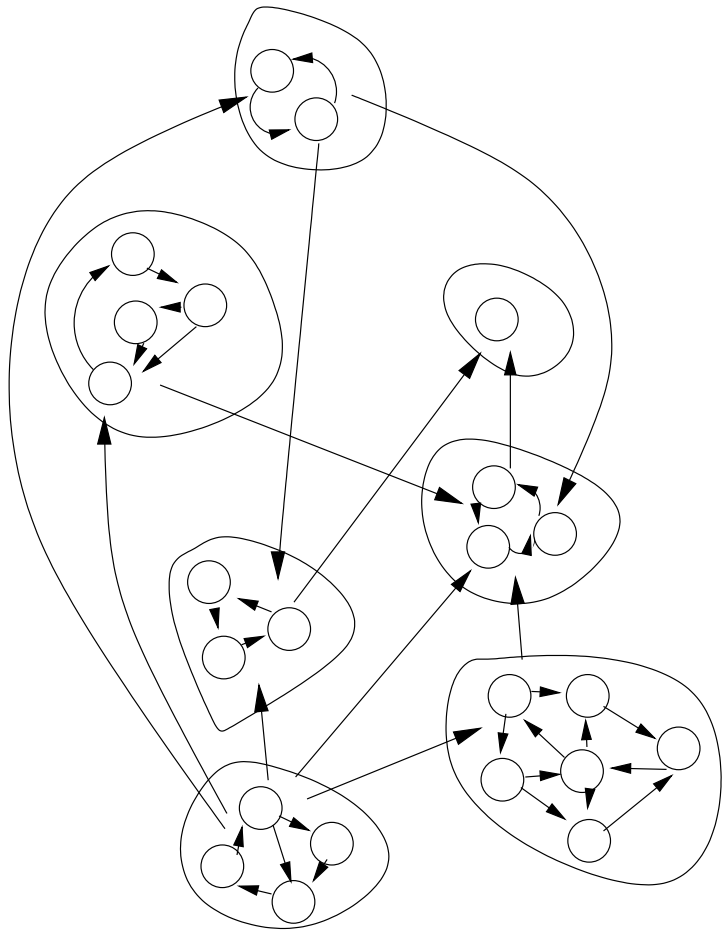}
\caption{The strongly connected components of a large system} 

\end{center}
\end{figure}

\paragraph*{Algorithm.}

\emph{Strongly\_connected\_components}($G$)

\begin{enumerate}
\item
  Use the Depth-First-Search (DFS) algorithm to compute $f[U]$ the 
  finishing time of $u$
\item 
  Compute $G^T=(V,E)$ where $E^T=\{(u,v) | (v,u)\in E\}$
\item
  Execute DFS on $G^T$ by grabbing vertices in the order of decreasing
  $f[u]$ as computed in step 1.
\item
  Output the vertices of each tree in the depth-first forest of step 3.
  as a separate strongly connected component
\end{enumerate}

\paragraph*{Complexity.}
This algorithm runs twice the time of DFS($G$) which is $\Theta(|V|+|E|)$

\subsection{Topological sort of graph}

If the system consists of a directed acyclic graph (DAG),
we can compute the topological sort of this graph
in order to have its hierarchical combination.

\paragraph*{Algorithm.}

\emph{Topological\_sort}($G$)

\begin{enumerate}
\item
  Call DFS($G$) to compute $f[u]$, the finishing time of $u$
\item 
  As each vertices is finished, put it into the front of a linked list
\item
  Return the linked list of vertices
\end{enumerate}

\begin{figure}
\begin{center}

\includegraphics[scale=0.8]{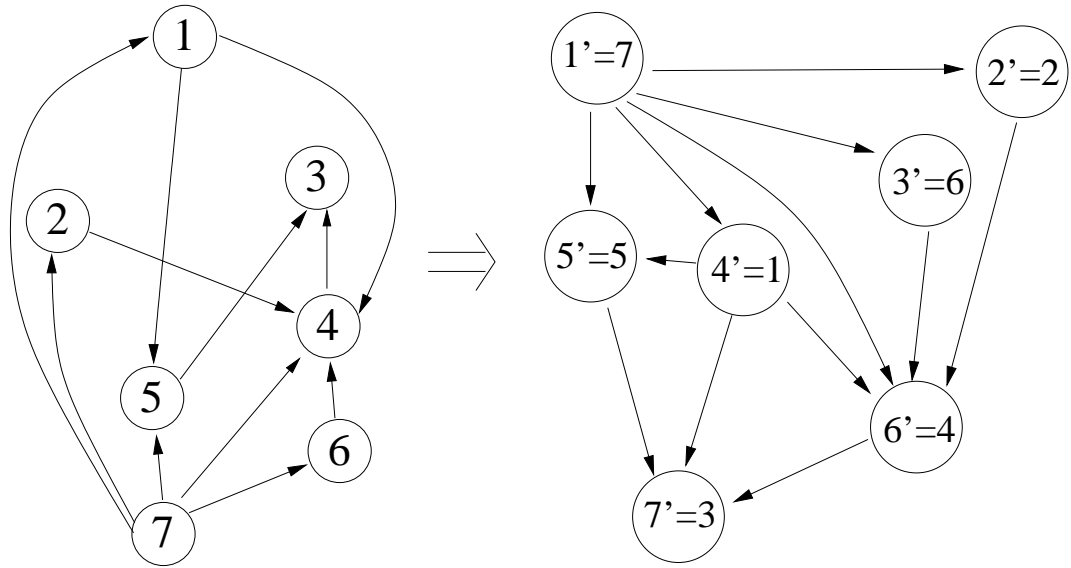}
\caption{The topological sort of the
  strongly connected components generated in figure 1.} 

\end{center}
\end{figure}

\paragraph*{Complexity.}
Since DFS($G$) takes $\Theta(|V| + |E|)$ and insertion into linked list 
cost $\theta(1)$ for each vertex, topological sort costs only 
$\Theta(|V| + |E|)$.

\subsection{Filtering large systems}

Once we have computed the \emph{strongly connected components}
of the large system $G$, we can consider the graph $G'=(V',E')$ consisting
of the strongly connected components of $G$ as vertices and  
$E'=\{(C_1,C_2)|\exists u \in C_1,v \in C_2 (u,v) \in E \}$.

\begin{prop}
$G'$ is a directed acyclic graph.
\end{prop}

Thus we can compute the topological sort of $G'$ which gives rise
to the hierarchical structure of the large system $G$.

Using the basic result on contraction of
hierarchies~\cite{Lohmiller98}, this implies that in order to show
that a large system is contracting, we only have to show that each
\emph{strongly connected component} of the system is contracting, and
that the couplings are bounded.

\section{Study of time varying hyper-curved attractors}

\subsection{Line attractor}

Consider a system ${\bf \dot{x}} = f({\bf x})$ contracting in a
constant metric ${\bf M}$. Then, the system 
$$
\left \{
\begin{array}{lcl}
  {\dot{s}} & = & 0 \\
  {\bf \dot{x}} & = & {\bf f}({\bf x}) - {\bf g}(s) 
\end{array}
\right.
$$
will be called a line attractor as ${\bf x}$ tends exponentially
towards ${\bf x_0}$ satisfying ${\bf f}({\bf x_0}) = {\bf g}(s)$. 

\subsection{Time varying hyper-curved attractor}

Consider a system ${\bf \dot{x}} = {\bf h}({\bf x},t)$ and suppose that there
exists an explicit metric in which the system can be rewritten :
$$
\left \{
\begin{array}{lcl}
  {\bf \dot{z}_1} & = & {\bf s}({\bf z_1},t)  \\
  {\bf \dot{z}_2} & = & {\bf f}({\bf z_2},t) + {\bf g}({\bf z_1},t)
\end{array}
\right.
$$
with $\frac{\partial {\bf f}}{\partial{\bf z_2}}({\bf z_2},t)$
uniformly negative definite.

Then the system is said to be a time varying hyper-curved attractor as
it tends to 
$$
{\bf z^\infty}({\bf z_1},t) =
\left (
\begin{array}{l}
  {\bf \alpha}({\bf z_1},t)  \\
  {\bf \beta}({\bf z_1},t)
\end{array}
\right)
$$

Define the virtual system
$$
  {\bf \dot{y}} = {\bf f}({\bf y},t) + {\bf g}({\bf z_1},t)
$$
This system is contracting as $\frac{\partial {\bf f}}{\partial{\bf y}}({\bf
  y},t) = \frac{\partial {\bf f}}{\partial{\bf z_2}}({\bf z_2},t)$ is
uniformly negative definite. 
So ${\bf y}$ tends exponentially to some ${\bf \beta}({\bf z_1},t)$.
As ${\bf z_2}(t)$ is another particular solution, we know from partial
contraction that ${\bf z_2}(t)$ tends to the same ${\bf \beta}({\bf z_1},t)$.
\qed

\paragraph*{Remark}

To know if hypothesis above are true given a system (with ${\bf h_1}$
and ${\bf h_2}$ assumed to be $\mathcal{C}^2$) 

$$
\frac{d}{dt} 
\left (
\begin{array}{l}
  {\bf z_1}  \\
  {\bf z_2}
\end{array}
\right)
=
\left (
\begin{array}{l}
  {\bf h_1} ({\bf z_1},{\bf z_2})  \\
  {\bf h_2} ({\bf z_1},{\bf z_2})
\end{array}
\right)
$$

we only have to check that

\begin{eqnarray*}
  \frac{\partial \ {\bf h_1}}{\partial {\bf z_2}} ({\bf z_1},{\bf z_2})
  =  0  \ \ \ \ \ \ \ \ \ \ \ \ \ \ \ \ \ \ \ \ \ \
\frac{\partial^2 \ {\bf h_2}}{\partial {\bf z_1} {\bf z_2}} ({\bf
  z_1},{\bf z_2}) & = & 0
\end{eqnarray*}

Then, from Schwarz theorem, we know that two other equations are true 
\begin{eqnarray*}
 \frac{\partial^2 \ {\bf h_1}}{\partial {\bf z_2} {\bf z_1}} ({\bf z_1},{\bf z_2})
  =  0  \ \ \ \ \ \ \ \ \ \ \ \ \ \ \ \ \ \ \
\frac{\partial^2 \ {\bf h_2}}{\partial {\bf z_2} {\bf z_1}} ({\bf
  z_1},{\bf z_2}) & = & 0
\end{eqnarray*}

and so the system can be rewritten in the required form.

\paragraph*{Example}
Consider the system
$$
\left \{
  \begin{array}{rcl}
    \dot s & = & \prod_{i=1}^n (s_i-s) \\
    \dot {\bf x} & = & - {\bf x} + {\bf f}(s)
  \end{array} 
\right.
$$ where the $s_i$'s are scalars.  This system is an attractor with
stable points $(s_i,{\bf f}(s_i))$.

\section{Composite variables}

\subsection{Estimation of the successive derivatives of a vector}
\label{sct:estimation} 
 
We show how to compute the $n$ successive derivatives of a given
vector only by assuming that the $(n+1)^{th}$ derivative of the vector
${\bf x}$ is zero.

Let $\widehat{\bf{x}}_i$ be the estimation of the $i^{th}$ derivative
of ${\bf x}$. We define each $\widehat{\bf{x}}_i$ associated composite
variable. 
$$
\widehat{\bf x}_i = \overline{\bf x}_i + \alpha_i {\bf x}
$$

and define the system :

$$
\dot{\overline{\bf X}} = {\bf J} \ {\widehat{\bf X}}
$$

With ${\bf X} = (x_1,\ldots,x_n)^T$ and

$$
{\bf J} =
\left(
\begin{array}{cccccc}
-\alpha_1 & 1 & . &   &  & . \\
. & 0 & 1 &  &  &  \\
. &   & 0 & 1 &   &  \\
. &   &   & 0 & 1 & . \\
. &   &   & . & 0 & 1 \\
-\alpha_n  & 0 &  &  & & 0
\end{array}
\right)
$$

Let us rewrite the system in ${\widehat{\bf X}}$:
$$
\dot{\widehat{\bf X}} = {\bf J} \ ({\widehat{\bf X} - {\bf Y}})
$$

with ${\bf Y} = (\dot{\bf x},0,\ldots,0)^T$ .
It is clear that the system is contracting iff the companion matrix
${\bf J}$ satisfies $\ {\bf J}^T\ M + M\ {\bf J} \ < \ 0\ $ for some metric $M$.

Now, assuming that the system is contracting, it is easy to see that
$\widehat{\bf X}_i = {\bf x}_i,\ \forall i$ is the unique solution of
the system. Indeed,
\begin{eqnarray*}
\dot{\widehat{\bf X}} & = & ({\bf x}_2,\ldots,{\bf x}_n,0)^T =  {\bf J} \ (0,{\bf x}_2,\ldots,{\bf x}_n)^T =  {\bf J} \ ({\widehat{\bf X} - {\bf Y}})
\end{eqnarray*}

So, if the system is contracting, we are sure to converge to the right
successive derivatives exponentially.

Note that we can check the contraction of the system above using the
Routh-Hurwitz criterion.

\subsection{Extension}
We can now analyze the case where the $(n+1)^{th}$ derivative of the
vector ${\bf x}$ is
a nonlinear function of ${\bf x},\dot{{\bf x}},\ldots$, that is 
${\bf x}_{n+1} = f({\bf x},{\bf x}_1,\ldots,{\bf x}_n)$.
We just have to replace 
$$
\dot{\overline{\bf X}} = {\bf J}\ {\widehat{\bf X}}
\quad \ \mathrm{by}
\quad \ 
\dot{\overline{\bf X}} = {\bf J}\ {\widehat{\bf X}} + (0,\ldots,0,f({\bf
  x},{\bf x}_1,\ldots,{\bf x}_n))^T 
$$
The Jacobian of the system is 
$$
{\bf J}_{ext} =
\left(
\begin{array}{cccccc}
-\alpha_1 & 1 & . &   &  & . \\
. & 0 & 1 &  &  &  \\
. &   & 0 & 1 &   &  \\
. &   &   & 0 & 1 & . \\
. &   &   & . & 0 & 1 \\
-\alpha_n + \frac{\partial f}{\partial \widehat{\bf x}_1} &
\frac{\partial f}{\partial \widehat{\bf x}_2} & . & . & . &
\frac{\partial f}{\partial \widehat{\bf x}_n} 
\end{array}
\right)
$$
Note that the result that $\alpha_n$ must be greater than $
\frac{\partial f}{\partial \widehat{\bf x}_n}$
was obtained in \cite{Slotine03} in the particular case where $f$
corresponds to a Van der Pol oscillator. In that case $\frac{\partial
  f}{\partial \widehat{\bf x}_i}$
were all null except for $\frac{\partial f}{\partial \widehat{\bf x}_n}
= 1 $,
and thus the condition was only $\alpha_n > 1$ 

\begin{figure}[h]
\centerline{\includegraphics[]{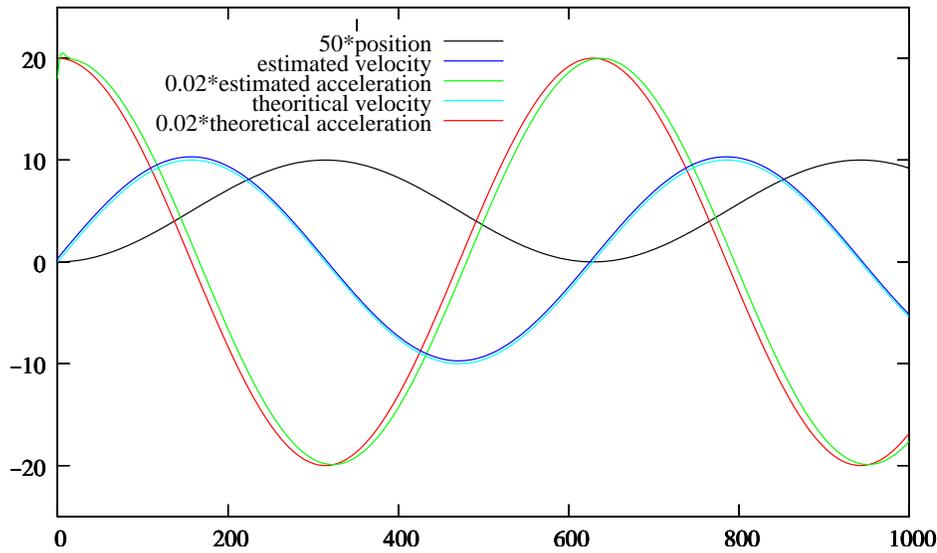}}
\caption{An estimation of a sinusoidal displacement}
 \label{fig:sinusoide}
\end{figure}

\paragraph*{Example}
We compute the velocity and
acceleration of the displacement $\frac{1-cos(100x)}{10}$ using the
differential equation $$\frac{d \ddot{x}}{dt} +5\ddot{x} + 10000\
\dot{x} = 0$$ The result can be seen in figure
\ref{fig:sinusoide} (we have shifted the estimated curves to
facilitate the analysis), where we have used the values $\alpha_1 =
5000$ and $\alpha_2 = 2000$.

\subsection{Estimation of velocity and acceleration using neural net}

\begin{figure}[h]
\centerline{\includegraphics[]{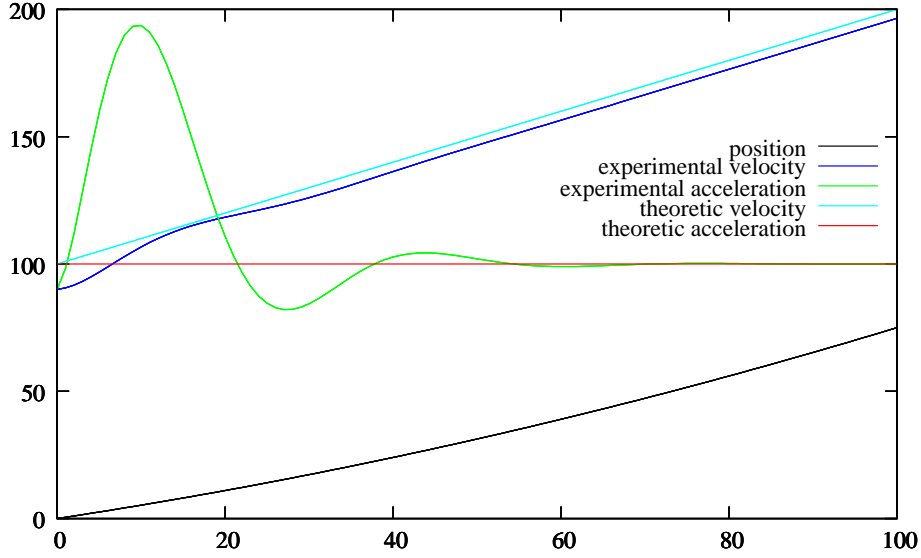}}
\caption{An estimation of velocity and acceleration of a parabolic trajectory}
 \label{fig:compositevariable}
\end{figure}

Composite variables can be used estimate the velocity and the
acceleration of a target given its position using a ``neural
network'', with potential application in modelling prediction.


As seen in (\ref{sct:estimation}), assuming that the acceleration of
the target is constant (ie. $\dot{\bf A} = 0$),  we can compute the
estimation of velocity and acceleration (resp. $\overline{\bf V}$ and
$\overline{\bf A }$) using only the position of the target $X$. For
that, we introduce two composite variables $\widehat{\bf V} =
\overline{\bf V} + \alpha {\bf X}$ and $\widehat{\bf A} =
\overline{\bf A} + \beta {\bf X}$ computed by the system : 
$$
\left\{
\begin{array}{l}
\dot{\overline{\bf V}} = -\alpha \ {\widehat{\bf V}} + \widehat{\bf A} =  
-\alpha \ {\overline{\bf V}} + (\beta-\alpha^2) {\bf X} + \overline{\bf A}\\
\dot{\overline{\bf A}} = -\beta \ {\widehat{\bf V}} = -
\beta \  \overline{\bf V} - \beta \ \alpha{\bf X} 
\end{array}
\right.
$$

We thus obtain a classical neural network :
$$
\tau \ \frac{d}{dt}  \left( 
\begin{array}{c}
  {\overline{\bf V}} \\
  {\overline{\bf A}} \\
  {\widehat{\bf V}} \\
  {\widehat{\bf A}}
\end{array}
\right) 
=
\left(
\begin{array}{cccc}
  -\tau \ \alpha & \tau & 0 & 0 \\ 
  -\tau \ \beta & 0  & 0 & 0\\
  1 & 0 & -1 & 0 \\
  0 & 1 & 0 & -1  
\end{array}
\right)
.
\left(
\begin{array}{c}
{\overline{\bf V}}\\
{\overline{\bf A}} \\
  {\widehat{\bf V}} \\
  {\widehat{\bf A}}
\end{array}
\right)
+
\left(
\begin{array}{c}
  -\tau \ (-\beta+\alpha^2) {\bf X}\\
  -\tau \ \beta \ \alpha {\bf X} \\
  \alpha {\bf X}\\
  \beta {\bf X}\\
\end{array}
\right)
$$

This network is contracting from section (\ref{sct:estimation})
and hierarchical analysis.

\paragraph*{Example}
A simulation of this system is presented in figure
\ref{fig:compositevariable}.
The small discrepancy between estimated and theoretical values is due
to the use of the ``neural network''.
\begin{figure}[h]
\centerline{\includegraphics[]{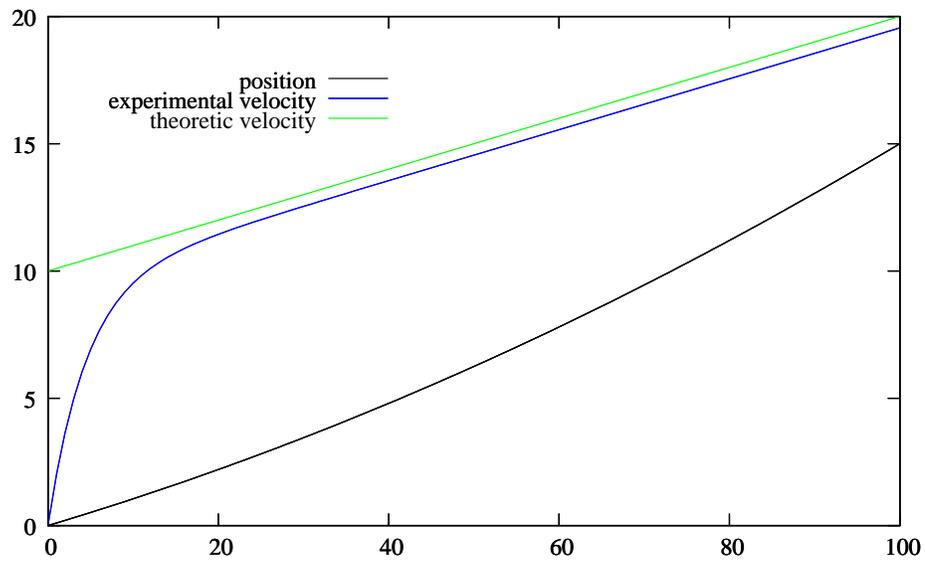}}
\caption{Converging beyond the scope of the system}
 \label{fig:compositevariable2}
\end{figure}

{\bf Acknowledgements:} This work was motivated by and benefited
greatly from an on-going collaboration with Alain Berthoz and Benoit
Girard at the College de France. NT was supported in part by a
European grant on the BIBA project.

\bibliographystyle{apalike}
\bibliography{biblio.bib}

\end{document}